\documentclass[aps,pra,twocolumn,superscriptaddress,
floatfix]{revtex4-1}
\usepackage{amsmath}
\usepackage[next]{inputenc}
\usepackage[dvips]{epsfig}
\usepackage{bbm,bm,bbold}
\usepackage{booktabs}
\usepackage{multirow}
\usepackage{hhline}
\usepackage{comment}
\usepackage{float}

\usepackage{amsmath,amsfonts,amssymb,amsthm}
\usepackage{color}
\usepackage{graphicx,latexsym}

\usepackage[colorlinks=true,urlcolor=blue,citecolor=blue,linkcolor=blue]{hyperref}

\usepackage{lipsum}

\usepackage{nameref}
\usepackage{varioref}
\usepackage{cleveref}

\usepackage{natbib}

\begin{document}
\renewcommand{\vec}{\mathbf}
\renewcommand{\Re}{\mathop{\mathrm{Re}}\nolimits}
\renewcommand{\Im}{\mathop{\mathrm{Im}}\nolimits}
\newcommand\scalemath[2]{\scalebox{#1}{\mbox{\ensuremath{\displaystyle #2}}}}

\title{Topological hybrid electron--hole Cooper pairing}

\author{Alexander Chansky}
\affiliation{School of Physics and Astronomy, Monash University, Victoria 3800, Australia}

\author{Dmitry K. Efimkin}
\email{dmitry.efimkin@monash.edu}
\affiliation{School of Physics and Astronomy, Monash University, Victoria 3800, Australia}
\affiliation{ARC Centre of Excellence in Future Low-Energy Electronics Technologies, Monash University, Victoria 3800, Australia}

\begin{abstract}
We consider electron--hole Cooper pair condensation in a heterostructure formed by a topological insulator film and a quantum well. We argue that the helical nature of the Dirac electronic states at the topological insulator surface results in the presence of two competing degenerate pairing channels. The corresponding paired states have an unconventional symmetry in the order parameter describing the Cooper pair condensate, can be classified by the topological Chern invariant, and are topologically distinct. We discuss possible manifestations of the nontrivial topology, including the formation of chiral states at the domain walls separating two distinct states, quantized anomalous transport phenomena, and connections with chiral topological superconductivity.   \end{abstract}

\date{\today}
\maketitle
\section{Introduction}
In recent years, there has been a growing interest in the topological states of matter, including topological insulators (Tis), semimetals, and superconductors~\cite{TopologicalInsulatorsReview,TopologicalSuperconductprsReview, WeylReview}. These materials host robust and unconventional surface electronic modes that have excellent prospects for energy-efficient electronic and spintronic devices. The Majorana modes in intrinsic or engineered topological superconducting setups~\cite{Majorana1,Majorana2} can potentially be applied as key elements for fault-tolerant quantum computation. 

Electron--hole Cooper pair condensates~\cite{LozovikYudson1,Shevchenko} are very close relatives of superconductors and can be driven by attractive Coulomb interactions in various closely spaced bilayer systems~\cite{EH1,EH2,EH3,EH4,EH5,EH7,Littlewood,GaAsRecent1,EHTI1,EHTI2,EHBG1,EHBG6,EHBGtwisted,Berman1,Berman2,Berman3} (see Ref.~\cite{EHreview} for a review). Until recently, condensation has been observed
only in the presence of a strong magnetic field that quenches the kinetic energies of electrons and holes and drastically enhances the effects of Coulomb interactions~\cite{EHQHEReview}. Some signatures of condensation~\footnote{However, it remains a matter of debate whether the quasi-long-range coherence and dipolar superfluidity that would potentially be useful for applications have been achieved~\cite{EfimkinAndreevDrag,EfimkinExcDrag,VignaleFluctDrag,AdamDrag}} have been reported via tunneling and Coulomb drag probes in double-bilayer graphene~\cite{JE1,JE2}, 
MoSe$_2$-WSe$_2$ heterostructures~\cite{BECExc1},  bilayers based on GaAs/AlGaAs quantum wells (QWs) ~\cite{DragQW1,DragQW2,DragQWRecent}, and the hybrid bilayer formed by GaAs/AlGaAs QW and graphene~\cite{Gamucci-Pellegrini-Natcomm-2014}.

An electron--hole Cooper pair condensate can also be topological if its order parameter has an unconventional symmetry (e.g., $p_x\pm i p_y$-wave states). Such states can be stabilized via interlayer tunneling in heterobilayers~\cite{TopEH1,TopEH2}, where the wave functions for electron and hole states are of different parities. However, interlayer tunneling also fixes the phase of the order parameter, and the resulting condensed state is a correlated TI rather than a superfluid. 

\begin{figure}[b]
    \centering
    \includegraphics[trim=4cm 11cm 4cm 1cm,scale=0.37]{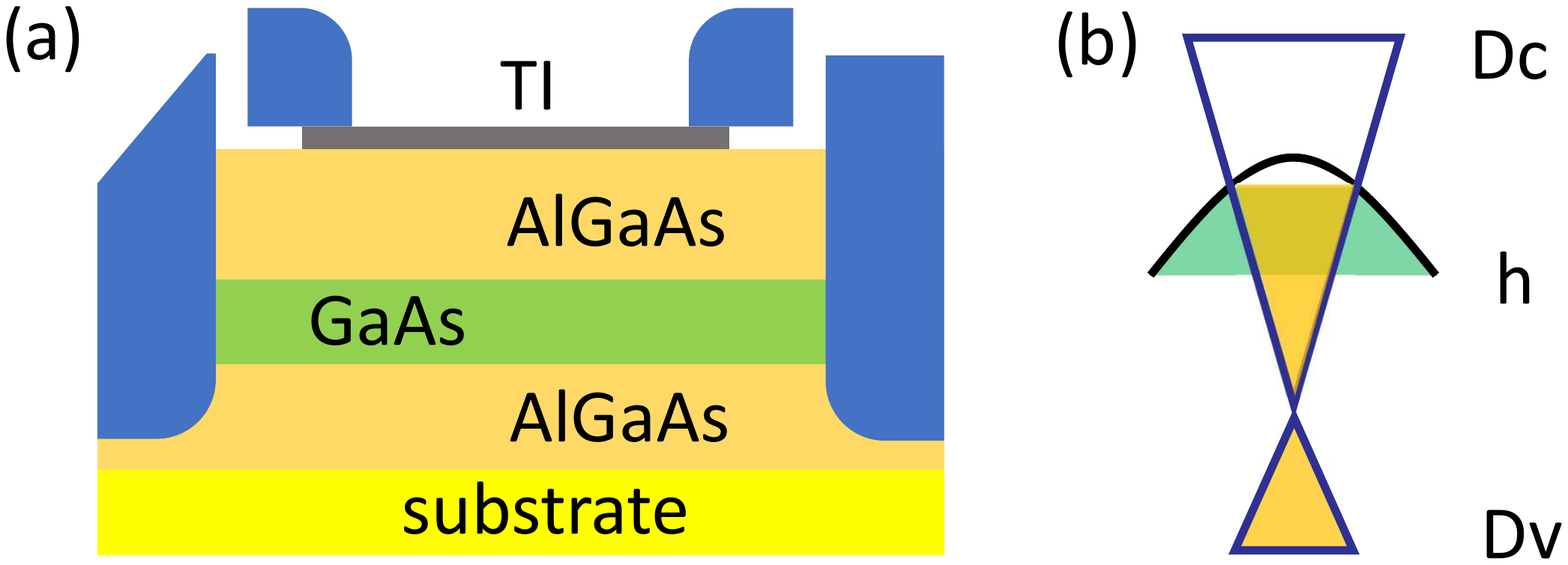}
    \caption{\label{fig:FigScheme} (a)
    Schematic of the heterostructure formed by a TI film and a GaAs QW. Carriers in the two layers are induced either by gating or by a doping layer (not shown in the sketch).  (b) The low-energy band structure of the system includes Dirac conduction (Dc) and valence (Dv) bands as well as a parabolic valance band (h) corresponding to hole-like states in the QW.}   \label{system}
\end{figure}

In this paper, we predict topological electron--hole Cooper condensate states that can be stabilized in the hybrid TI-QW heterostructure sketched in Fig.~\ref{fig:FigScheme}-(a). We argue that the helical nature of the Dirac electronic states at the TI surface results in the presence of two competing degenerate Cooper pairing channels. The corresponding paired states have an unconventional symmetry of the order parameter, can be classified by the topological Chern invariant, and are topologically distinct. We discuss possible manifestations of nontrivial topologies, including the formation of chiral states at the domain walls separating two distinct states, quantized anomalous transport phenomena, and connections with chiral topological superconductivity.

.  
\section{Model}
The hybrid bilayer formed by the TI-QW heterostructure and the corresponding arrangement of the electronic bands are sketched in Fig.~\ref{fig:FigScheme}-(a) and (b). In this paper, we develop a minimal generalization of the Bardeen--Cooper--Schrieffer (BCS) theory that properly accounts for the helical nature of Dirac electrons at the TI surface. Because the Dirac electronic spectrum is gapless and does not favor the formation of excitons by electrons and holes, this approach properly describes the phenomenology of condensation in the weak- to moderate-coupling regime (see Appendix A for an extended discussion).

The Dirac surface states in the TI have a spin degree of freedom and are described by the field operator $\hat{\psi}_\mathrm{D}=\{\hat{\psi}_{\uparrow},\hat{\psi}_{\downarrow}\}$. Their kinetic energy is given by 
\begin{equation}
\hat{H}_{\mathrm{D}}=\int d^2 r \; \psi^\dagger_\mathrm{D}(\vec{r})(v_\mathrm{e} [\bm{\sigma}\times  \vec{p}]_z-v_\mathrm{e} p_\mathrm{F}) \psi_\mathrm{D}(\vec{r}).
\end{equation}
The Fermi level is assumed to be in the Dirac conduction band, and the corresponding dispersion of electronic states is linear, $\xi_{\vec{p}}^{\mathrm{e}}=v_{\mathrm{e}}(p-p_\mathrm{F})$, with velocity $v_\mathrm{e}$ and Fermi momentum $p_\mathrm{F}$. The helical nature of the Dirac electrons is encoded in their spinor wave function, $|\vec{p}\rangle =\{  e^{- i \phi_\vec{p}/2}, i e^{i \phi_\vec{p}/2}\}^{\mathrm{T}}/\sqrt{2}$, where $\phi_\vec{p}$ is the polar angle for the electron momentum $\vec{p}$.  

In the adjacent QW, there is a deficit of hole-like states. These states follow a conventional parabolic dispersion law, $\varepsilon^\mathrm{h}_\vec{p}=-\vec{p}^2/2 m_\mathrm{h}$, with mass $m_\mathrm{h}$ and can be described by the field operator $\psi_\mathrm{S}$. Their kinetic energy can be presented as  
\begin{equation}
H_\mathrm{S}=\int d^2r \psi^\dagger_\mathrm{h}(\vec{r})\left(\frac{p^2_\mathrm{F}}{2m_\mathrm{h}}-\frac{\vec{p}^2}{2m_\mathrm{h}}\right)\psi_\mathrm{h}(\vec{r}).
\end{equation}
The Fermi momentum for the holes is assumed to match the electronic momentum $p_\mathrm{F}$ that gives the most favorable condition for electron--hole Cooper pairing.  Due to the spin degeneracy of holes, only one of two species participates in the Cooper pairing and is explicitly considered, while the other species remains intact.  

Cooper pairing between electrons and holes is driven by the attractive Coulomb interactions between them, which can be approximated by the contact pseudopotential $V$~\footnote{It should be noted that we refer to the empty states
in the valence band as holes, but do not perform the formal particle-hole transformation. As a result, the attractive interactions correspond to $V>0$.} as 
\begin{equation}
H_{\mathrm{int}}=\int d^2r V \psi^\dagger_{\mathrm{D}}(\vec{r}) \psi^\dagger_{\mathrm{h}}(\vec{r}) \psi_{\mathrm{h}}(\vec{r}) \psi_\mathrm{D}(\vec{r}). 
\end{equation}
We omit the intralayer repulsive interactions because their main effect results in a renormalization of the dispersion relations for electrons and holes.

Before presenting the BCS-like mean-field theory for the paired state, it is instructive to discuss the corresponding Cooper pair problem. 

\section{Cooper pair problem}
The formation of an electron--hole bound state with the restriction of using only states outside the Fermi seas can be described by the following eigenvalue problem:
\begin{equation}
\label{EqCooperPairProblem}
(\xi^\mathrm{e}_\vec{p}-\xi^\mathrm{h}_\vec{p})C_\vec{p}- \sum_{|\vec{p}'|>p_\mathrm{F}}  V \Lambda_{\vec{p},\vec{p}'}C_{\vec{p}'}=E C_{\vec{p}}.     
\end{equation}
The linear slope of the Dirac dispersion is of little importance because the quadratic curve for holes can also be approximated as linear in the vicinity of the Fermi level $\xi^\mathrm{h}_\vec{p}=-v_\mathrm{h}(p-p_\mathrm{F})$ with $v_\mathrm{h}=p_\mathrm{F}/m_\mathrm{h}$. In contrast, the helical nature of Dirac states plays a critical role, is manifested via the overlap of the spinor wave functions:  
\begin{equation}
\label{Lambda}
\Lambda_{\vec{p}\vec{p}'}=\langle \mathrm{c} \vec{p}| \mathrm{c}\vec{p}' \rangle=\cos\left(\frac{\phi_{\vec{p}\vec{p}'}}{2}\right) 
\end{equation}
and shapes the Cooper pairing channels. In the conventional case (e.g., QW-QW bilayer with $\Lambda_{\vec{p}\vec{p}'}=1$), contact interactions drive Cooper pairing only in the s-wave channel. In the hybrid case, there are two degenerate competing channels with angular numbers $l=\pm 1/2$. The corresponding binding energies are equal to each other and are given by $E_{\mathrm{CP}}=2 v_0 p_0 \exp[-2/\lambda]$. Here, $\lambda=\nu_\mathrm{F}V/2$ is the coupling constant, $v_0=(v_\mathrm{e}+v_\mathrm{h})/2$ is the average velocity, $\nu_\mathrm{F}=p_\mathrm{F}/2 \pi\hbar^2 v_0$ is the effective density of states at the Fermi level, and $p_0\ll p_\mathrm{F}$ is the momentum cutoff for pairing correlations. 


The presence of the two pairing channels hints a competition between the two distinct paired states, which can be described using the generalized BCS theory.

\section{Mean-field theory}
\subsection{Mean-field theory and order parameters} 
In the mean-field theory, the Hamiltonian of the system decouples as 
\begin{equation}
\label{MeanFieldH}
    H=\int{d^2r \left[ \hat{\Psi}^\dagger_\mathrm{Dh} H^{\mathrm{Dh}}_\mathrm{BdG}
    \hat{\Psi}_\mathrm{Dh}
    - \frac{2(|\Delta_\uparrow|^2+|\Delta_\downarrow|^2)}{V}\right]}.
\end{equation}
Here, $\hat{\Psi}_\mathrm{Dh}=\{\hat{\psi}_\mathrm{D},\psi_\mathrm{h}\}$, and $\hat{H}^{\mathrm{Dh}}_\mathrm{BdG}$ is the Bogoliubov--de Gennes (BdG) Hamiltonian given by
\begin{equation}
\label{BdG3x3}
    \hat{H}^{\mathrm{Dh}}_{\mathrm{BdG}}= 
    \begin{pmatrix}
    v_\mathrm{e} [\bm{\sigma}\times  \vec{p}]_z-v_\mathrm{e} p_\mathrm{F} & \sqrt{2} \hat{\Delta} \\
    \sqrt{2} \hat{\Delta}^\dagger & \frac{p^2_\mathrm{F}}{2m_\mathrm{h}}-\frac{\vec{p}^2}{2m_\mathrm{h}} 
    \end{pmatrix}.
\end{equation}
The order parameter for the Cooper pair condensate has two components,  $\hat{\Delta}=\{\Delta_\uparrow, \Delta_\downarrow \}$, which describe the selective pairing correlations with Dirac electrons with different spin projections as $\hat{\Delta}(\vec{r})=-V\langle \psi_{\mathrm{S}}^\dag(\vec{r})\hat{\psi}_{\mathrm{D}}(\vec{r})\rangle/\sqrt{2}$. In the considered weak- to moderate-coupling regime, $|\Delta_{\uparrow(\downarrow)}|\ll v_\mathrm{e} p_\mathrm{F}, v_\mathrm{h} p_\mathrm{F}$, the pairing correlations do not extend to the remote Dirac valence band and this band is of minimal importance. It is instructive to perform a unitary rotation to the Dirac band basis and truncate the valence band, i.e., $\hat{H}^{\mathrm{Dh}}_\mathrm{BdG}\rightarrow \hat{H}^{\mathrm{eh}}_\mathrm{BdG}$ and $\hat{\Psi}_\mathrm{Dh}\rightarrow \hat{\Psi}_\mathrm{eh}$. The resulting BdG Hamiltonian is given by 
\begin{equation}
\label{BdG2x2}
\hat{H}^{\mathrm{eh}}_{\mathrm{BdG}}= 
    \begin{pmatrix}
    \xi_{\vec{p}}^\mathrm{e} & \Delta_{\vec{p}} \\
    \Delta_{\vec{p}}^* & \xi_{\vec{p}}^\mathrm{h}
    \end{pmatrix}.
\end{equation}
As hinted by the Cooper pair problem, the order parameter \begin{equation}
\label{DeltaProjected}
\Delta_\vec{p} =e^{i\frac{\phi_\vec{p}}{2}} \Delta_{\uparrow}-i   e^{-i\frac{\phi_\vec{p}}{2}} \Delta_{\downarrow}
\end{equation}
has two angular harmonics with $l=\pm 1/2$. These two channels correspond to the selective spin-resolved pairing correlations, and the degeneracy between them is protected by time-reversal symmetry.

The Cooper pairing mixes electron and hole states from different layers. The dispersions for the resulting  Bogoliubov quasiparticles are given by
\begin{equation} 
\varepsilon_{\vec{p}}^{\pm}=\frac{\xi^\mathrm{e}_\vec{p}+ \xi^\mathrm{h}_\vec{p}}{2 }\pm \sqrt{\left(\frac{\xi^\mathrm{e}_\vec{p}- \xi^\mathrm{h}_\vec{p}}{2 }\right)^2+|\Delta_\vec{p}|^2},
\end{equation}
where $|\Delta_\vec{p}|$ is given by 
\begin{equation*}
|\Delta_\vec{p}|^2=  |\Delta_\uparrow|^2+  |\Delta_\downarrow|^2-  2  |\Delta_\uparrow| |\Delta_\downarrow| \sin(\phi_\vec{p}-\Phi).
\end{equation*}
Here, $\Phi$ is the relative phase for two order parameters. 
Due to the asymmetry of the dispersion relations, $v_\mathrm{e}\neq v_\mathrm{h}$, the extrema for the dispersion curves $
\varepsilon_{\vec{p}}^{\pm}$ are shifted from the Fermi momentum $p_\mathrm{F}$, and the resulting gap is indirect and reduced to $2\Delta_\vec{p}^\mathrm{g}$, where $\Delta_\vec{p}^\mathrm{g}=|\Delta_\vec{p}| \sqrt{v_\mathrm{e} v_\mathrm{h}/v_0^2}$.

In general, the angular profile of the gap is asymmetric, and its orientation is determined by the relative phase $\Phi$. The magnitude of $|\Delta_\vec{p}|$ smoothly changes between the values of $||\Delta_+|\pm|\Delta_-||$ achieved at $\phi_\vec{p}=\Phi\pm \pi/2$. As a result, the state with the coexisting order parameters spontaneously breaks the rotational symmetry and can therefore be interpreted as a nematic electron--hole Cooper pair condensate. However, if either of the two order parameters vanishes, the rotational symmetry is restored.

The self-consistency equation for the order parameter
\begin{equation}
\Delta_\vec{p}=\sum_{\vec{p}'}  V \Lambda_{\vec{p},\vec{p'}}\frac{n_\mathrm{F}(\varepsilon_{\vec{p}'}^{-})-n_\mathrm{F}(\varepsilon_{\vec{p}'}^{+})}{\varepsilon_{\vec{p}'}^+-\varepsilon_{\vec{p}'}^-} \Delta_{\vec{p}'}  
\end{equation}
includes the same factor $\Lambda_{\vec{p}\vec{p}'}$ as the Cooper pair problem, Eq.~(\ref{EqCooperPairProblem}). 
At zero temperature, two solutions, $\hat{\Delta}=\{ \Delta_0,0\}$ and $\hat{\Delta}=\{ 0, \Delta_0\}$ with $\Delta_0=2 v_0 p_0 \exp[-1/\lambda]$, can be readily identified and correspond to two distinct isotropic paired states. To rule out the nematic paired states, it is instructive to address the energetics of the Cooper pairing in more detail.  

\begin{figure}
    \centering
    \includegraphics[scale=1.0]{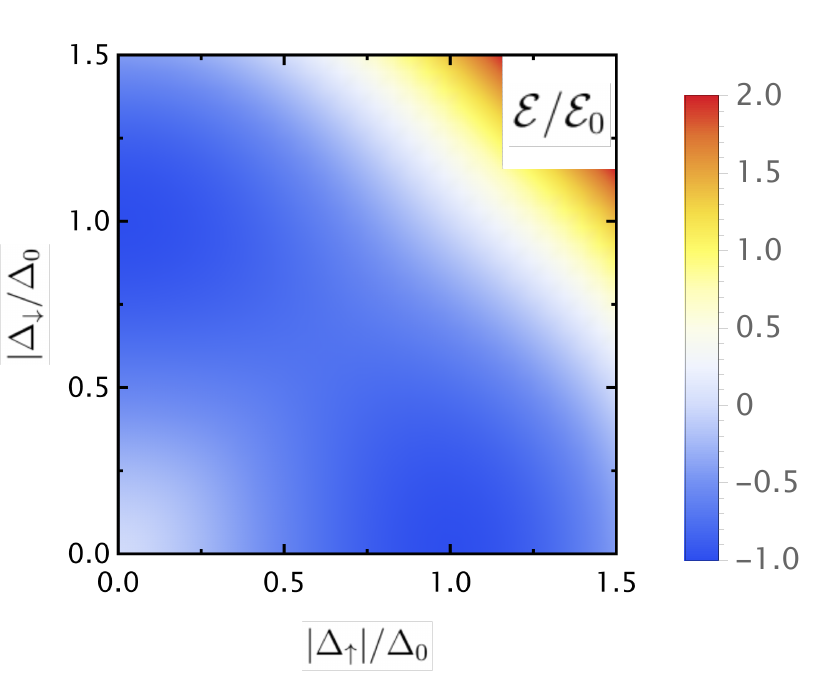}
\caption{Dependence of the condensation energy $\mathcal E$ on the magnitude of the order parameters $|\Delta_{\uparrow(\downarrow)}|$. As we discuss in the main text, there are two degenerate minima corresponding to the solutions $\hat{\Delta}=\{ \Delta_0,0\}$ and $\hat{\Delta}=\{ 0, \Delta_0\}$, while the coexistence of the two order parameters is energetically unfavorable.}
\label{condenergy}
\end{figure}

\subsection{Energetics}
The condensation energy $\mathcal E$ is given by the difference between the ground-state energies of the paired and normal states, as evaluated in Appendix B. This energy depends only on the magnitude of the order parameters,  $|\Delta_{\uparrow (\downarrow)}|$, not on their relative phase, $\Phi$ (the phase $\Phi$ determines only the orientation of the angular profile for the order parameter $\Delta_\vec{p}$). The corresponding dependence is presented in Fig.~\ref{condenergy} and shows two minima of the same depth. For each minimum, one of the two order parameter amplitudes vanishes, which implies that the nematic electron--hole Cooper pair condensation is  energetically unfavorable. This behavior is also encoded in the Ginzburg--Landau (GL) functional for the free energy density, which is  evaluated in Appendix C and is given by 
\begin{eqnarray}
\mathcal F_\mathrm{GL}(|\Delta_\uparrow|,|\Delta_\downarrow |)=a ( |\Delta_\uparrow|^2+|\Delta_\downarrow|^2)\nonumber\\
    +\frac{b}{2}\left(|\Delta_\uparrow|^2+|\Delta_\downarrow|^2\right)^2 + 2 b|\Delta_\uparrow|^2|\Delta_\downarrow|^2.
\end{eqnarray}
 Here, $b>0$ and $a$ switches its sign at the critical temperature of condensation. The last term describes the repulsive interactions between the order parameters, which inhibit their coexistence.  
\subsection{Estimations}
For the massless surface Dirac states, we choose $\epsilon^\mathrm{e}_\mathrm{F}\approx50\;\hbox{meV}$ and $v_\mathrm{e}\approx 6.7 \times \; 10^7\; \hbox{cm}/\hbox{s}$, which implies that the electron concentration is $n_\mathrm{e}=10^{11}\;\hbox{cm}^{-2}$. For the hole-like states in the QW,  we choose an effective of mass $m_\mathrm{h}=0.09\; m_0$, where $m_0$ is electron mass in free space, which implies that the Fermi energy $\epsilon^\mathrm{h}_\mathrm{F}\approx4.7\;\hbox{meV}$, concentration $n_\mathrm{h}=2 \times 10^{11}\;\hbox{cm}^{-2}$ and the Fermi velocity $v_\mathrm{h}=1.4 \times 10^7\; \hbox{cm}/\hbox{s}$. The reduction of the gap in the Bogoliubov quasiparticle spectrum due to the asymmetry of the electron and hole dispersion relations is given by the factor $\Delta^{\mathrm{g}}_{\vec{p}}/|\Delta_{\vec{p}}|=\sqrt{v_\mathrm{e} v_\mathrm{h}/v_0^2}\approx 0.75$ and is very small. We also choose $p_0\approx 0.4 p_0 $ and use the static screening approximation (which has been argued to underestimate the critical temperature of Cooper pair condensation~\cite{EHMultiband1,EHMultiband2,EHMultiband3}) to estimate the maximal coupling constant as $\lambda\approx 1/6$. The resulting temperature $T_0\approx 0.3\;\hbox{K}$ is comparable to the temperature range within which anomalous interlayer tunneling and Coulomb drag effects have been reported in other electron--hole setups. However, it is smaller compared to the predictions for the moderate- to strong-pairing regime, which can be engineered in certain closely spaced bilayer systems, including the ones composed of graphene bilayers~\cite{EHBG1}, transition metal dichalcogenide monolayers~\cite{Berman2}, and phosphorene~\cite{EH7, Berman3}. A reliable calculation of the critical temperature requires advanced approximations for the effects of Coulomb interaction screening,  disorder, and long-range density variations that act against electron--hole condensation and is beyond the scope of this paper.

\subsection{Nontrivial topology and domain wall states}
The phase winding in the angular dependence of the order parameter $\Delta_\vec{p}$ given by Eq.~(\ref{DeltaProjected}) hints at a nontrivial topology for the spectrum of the Bogoliubov quasiparticles. In the paired state $\Delta_\uparrow$, two branches   $\varepsilon_{\vec{p}}^{\pm}$ have the Chern numbers
$\mathcal{C}^\uparrow_{\pm}=\pm 1/2$, whereas in the state $\Delta_\downarrow$, the values swap, $\mathcal{C}^\downarrow_{\pm}=\mp 1/2$. Thus, the two competing paired states can be interpreted as topologically nontrivial and distinct. 

Typically, nontrivial topology manifests itself via unconventional and protected edge states. However, the presence of these states also requires a global energy gap at both sides of the edge. This is not the case for the heterostructure sketched in Fig.~\ref{fig:FigScheme}-(a), which terminates with the TI film. Outside the film, only hole-like states are present, and their gapless spectrum does not favor any edge states. Instead, the protected modes are anticipated to appear at the domain walls separating two topologically distinct paired states, $\Delta_\uparrow$ and $\Delta_\downarrow$. The Chern number difference across such a domain wall is equal to unity, and the presence of a single chiral state is dictated by the topology.

\begin{figure}
    \centering
    \includegraphics[scale=1.0]{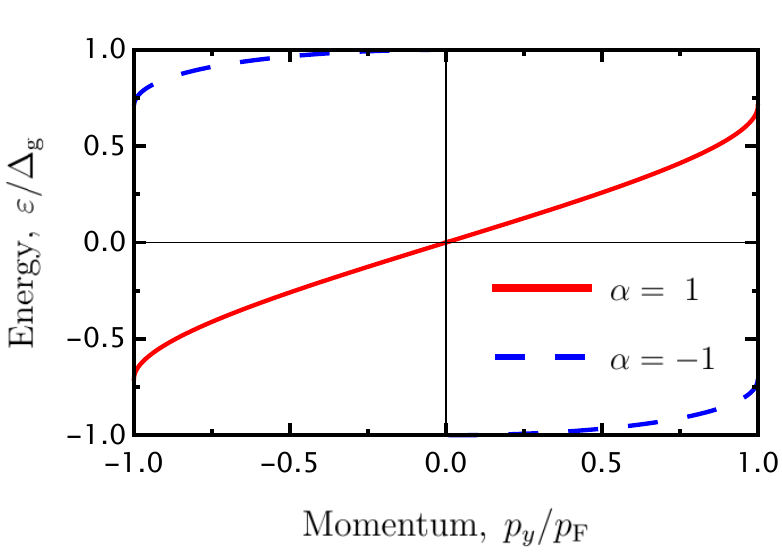}
\caption{Spectrum of states hosted by the domain wall separating regions with distinct paired states, $\Delta_{\uparrow(\downarrow)}$. Consistent with the topological analysis, there is a single chiral state that tends to cross the energy gap. The unexpected discontinuities at $p_y=\pm p_\mathrm{F}$ are artifacts of the Andreev approximation, which is valid at $p_y\ll p_\mathrm{F}$.}
\label{DWspectrum}
\end{figure}

An accurate calculation of the domain wall states is mathematically challenging because the spatial profile of the order parameter $\hat{\Delta}(\vec{r})$ and the spectrum of the BdG equations must be evaluated self-consistently~\footnote{It should be mentioned that this also relevant for the domain walls in chiral superconductors separating states with $p_x\pm i p_y$-wave symmetry of the order parameter.}. However, the presence of a single chiral state can be demonstrated by a simplified approach that has been elaborated to describe the spectrum of domain wall states in topological chiral superconductors~\cite{DWpwave1,DWpwave2}. First, we assume that the domain wall has a step-like profile, $\hat{\Delta}(x)=\Theta (-x)\{0,\Delta_0  \}+ \Theta (x)\{\Delta_0 e^{i \Phi}, 0 \}$, where $\Theta(x)$ is the Heaviside function and $\Phi$ is the phase difference for the order parameters across the domain wall. Second, we use the Andreev (or semiclassical) approximation that utilizes $\Delta_\vec{p}\ll v_{\mathrm{e}(\mathrm{h})} p_\mathrm{F}$ and is based on the following \emph{ansatz} for the wave function:
\begin{equation}
\hat{\Psi}_\mathrm{eh}(x,y)=\sum_{\alpha=\pm 1} \hat{\chi}_\alpha(x) e^{i\left(\alpha \sqrt{p_\mathrm{F}^2-p_y^2} x + p_\mathrm{y}y\right)}. 
\end{equation}
Here, $\chi_\alpha(x)$ is a slowly varying envelope function, and $\alpha=\pm 1$ labels two points at the Fermi line that have the same momentum $p_y$ along the domain wall. These points are assumed to be very well separated; thus, the Andreev approximation is reliable only at $p_y\ll p_\mathrm{F}$. As demonstrated in Appendix D, $\chi_\alpha(x)$ satisfies the Dirac-like equation, and the dispersion of states hosted by the domain wall is given by
\begin{equation}
\varepsilon_{p_y}^{\alpha n} =\Delta_\mathrm{g} \sin \left[\frac{\Phi}{2\alpha}-\frac{\pi}{4} + \frac{1}{2}\arcsin\left(\frac{p_y}{p_\mathrm{F}}\right) +\pi n\right].
\end{equation}
Here, $n$ is an integer, and $2\Delta_\mathrm{g}=2\Delta_0 \sqrt{v_e v_h/v_0^2}$ is the gap in the spectrum of Bogoliubov quasiparticles~\footnote{Only the solutions with a sine argument within $-\pi/2$ and $\pi/2$  are physically relevant.}. Unless $\Phi=\pm \pi/2$, the dispersion relation $\varepsilon_\alpha (p_y)$ has discontinuities at the intermediate momentum $-p_\mathrm{F}<p_\mathrm{y}<p_\mathrm{F}$. The corresponding \emph{ansatz} do not capture the actual spatial profile of the order parameter $\hat{\Delta}(x)$ and can therefore be safely ruled out~\cite{DWpwave1,DWpwave2}. The dispersion relation of the domain wall states at $\Phi=\pi/2$ is presented in Fig.~\ref{DWspectrum}, whereas the results for $\Phi=-\pi/2$ can be obtained by switching $\alpha\rightarrow-\alpha$. Consistent with the topological analysis, the domain wall hosts a single chiral state that tends to cross the energy gap. The unexpected discontinuities at $\pm p_\mathrm{F}$ are artifacts of the Andreev approximation, while the dispersion of chiral states in the domain of its validity can be approximated as linear,  $\varepsilon_{p_y}\approx \Delta_\mathrm{g}p_y/2p_\mathrm{F}$. The state propagates in only one direction and is, therefore, immune to backscattering. 

\section{Discussion}
Probing the nontrivial topologies of electron--hole Cooper pair condensates does not require separate contacts for electrons and holes (which are essential for dipolar superfluid phenomena). In this setup, two layers are connected in parallel, and the condensate flow does not contribute to the electric current through the system. Instead, the transport phenomena are solely determined by the Bogoliubov quasiparticles, and the magnitude of the Hall conductivity,  $\sigma_\mathrm{H}=C^{\uparrow(\downarrow)}_{-}e^2/2\pi \hbar$, in the state $\Delta_{\uparrow(\downarrow)}$ is given by the Chern number for the lower branch. The paired state, however, is not the anomalous quantum Hall insulator phase because one of two species of the hole-like states remains unpaired. This behavior can be confirmed in transport experiments. 

The pair of topologically distinct electron--hole paired states is reminiscent of the topological chiral superconducting states with $p_x\pm i p_y$-wave symmetry. However, there are a number of subtle differences between these states. First, the phase winding of the order parameters $\Delta_{\uparrow(\downarrow)}$ across the closed loop in momentum space is equal to $\pi$, rather than $2\pi$, as it is in chiral superconductors. As a result, the topological numbers for the electron--hole paired states are half-integer, whereas the topological numbers for the superconducting $p_x\pm i p_y$-wave states are integers, $\mathcal C=\pm 1$ . Second, pairing electrons in the chiral superconductor are indistinguishable (they are effectively spinless), which is not the case for the pairing of electrons and holes. Third, the hybrid paired states do not respect particle--hole symmetry unless $v_{\mathrm{e}}=v_{\mathrm{h}}$, which is not the case in realistic conditions. Symmetry is an essential ingredient for protecting zero-energy Majorana states localized in a vortex in chiral superconductors. 

The degeneracy of pairing channels relies on time-reversal symmetry. In the presence of an out-of-plane magnetically induced exchange field $\delta$, which plays the role of the Dirac mass and can be either intrinsic (magnetic TIs~\cite{MagneticTI1,MagneticTI2,MagneticTI3,MagneticTI4}) or extrinsic (magnetic doping~\cite{MagneticDoping1,MagneticDoping2,MagneticDoping3}, or the proximity effect~\cite{ProximityReview1,ProximityReview2}), the coupling constants are modified as $\lambda_{\uparrow(\downarrow)}=(1\pm\delta/ \sqrt{v_\mathrm{e}^2p_\mathrm{F}^2+\delta^2})\nu_\mathrm{F}V/2$ (see Appendix E for an extended discussion). Due to the exponential dependence of Cooper pairing on the coupling constants, even the presence of a small gap $|\delta|\ll v_\mathrm{e}p_\mathrm{F}$ has a profound effect. This result opens interesting opportunities not only to magnetically control the equilibrium paired state, but to engineer domain walls and vortices in the profile of the order parameter by using spatially varying magnetic textures, i.e., magnetic domain walls, vortices~\cite{MagneticVorticies}, merons~\cite{MagneticMerons}, or skyrmions~\cite{MagneticSkyrmions}. 

The presence of the magnetic order drastically increases the condensation energy due to its exponential sensitivity to the leading coupling constant. As a result, the magnetic order can be driven by Cooper pair condensation and co-facilitated by intralayer repulsion, which is not sufficiently strong to produce the ordering alone~\cite{TIMagneticInstability}. This mechanism is reminiscent of the interplay between superconductivity and loop-current correlations on a honeycomb lattice~\cite{LoopCurrents}. 

The spontaneous formation of Cooper pair condensates in electron--hole bilayers modifies the spectra of plasmonic excitations~\cite{EHPlasmon1} and results in an additional collective mode with a Higgs-like nature~\cite{HiggsMode}. The presence of two (almost) degenerate pairing channels in the TI-QW bilayer favors Bardasis--Schrieffer modes~\cite{BS0,BS2,BS1}. These modes correspond to the propagating oscillations of $\Delta_{\uparrow(\downarrow)}$ in the pair with the order parameter $\Delta_{\downarrow (\uparrow)}$ and can be probed in both far- and near-field optical experiments~\cite{BS1}.   

Interestingly, the presence of multiple degenerate pairing channels is not a unique feature of the TI-QW bilayer. As we argue in Appendix F, this phenomenon is generic for hybrid electron--hole bilayers, e.g., for bilayers formed by a few-layer graphene sheets of different chiralities~\cite{EH5}. In contrast, for all corresponding homobilayers, the s-wave pairing channel is dominant.

To conclude, we have predicted topological electron--hole Cooper condensate states in a hybrid TI-QW heterostructure and have discussed possible manifestations of nontrivial topologies, including the formation of chiral states at domain walls separating two distinct states, quantized anomalous transport phenomena, and connections with chiral topological superconductivity. 

\acknowledgments 
We acknowledge useful discussions with Alex Hamilton, David Neilson and Hong Liu.  We acknowledge support from the Australian Research Council Centre of Excellence in Future Low-Energy Electronics Technologies (CE170100039).

\appendix

\section{Weak- to moderate-pairing regime}
The nature of correlations in the considered electron--hole bilayer is characterized by three dimensionless parameters. The first two parameters scale the ratio of repulsive interactions and kinetic energy in two layers and are given by $r^{\mathrm{h}}_\mathrm{s}=m_\mathrm{e} e^2/\hbar p_\mathrm{F} \kappa_{\mathrm{h}}$ and $r^{\mathrm{e}}_\mathrm{s}=e^2/\hbar v_\mathrm{e} \kappa_{\mathrm{e}}$, respectively.  Here, $\kappa_\mathrm{e}$ and $\kappa_\mathrm{h}$ are the effective dielectric constants for each layer. The third parameter, $p_\mathrm{F}d_\mathrm{eh}/\hbar$, scales the interlayer Coulomb attraction with respect to the intralayer repulsion, where $d_\mathrm{eh}$ is the interlayer distance.  If $r_\mathrm{s}^{\mathrm{e}(\mathrm{h})}\ll1$ and $p_\mathrm{F}d_\mathrm{eh}/\hbar\gg1$, the system is in the weak-coupling regime. In this regime, the pairing correlations are prominent only in the vicinity of the Fermi level for both electrons and holes and it can be described by the BCS theory. Its range of the applicability in the considered TI-QW heterostructure also extends to the moderate-coupling regime, $r_\mathrm{s}^{\mathrm{e}(\mathrm{h})}\sim1 $ and $p_\mathrm{F}d_\mathrm{eh}/\hbar\sim1$, and is considerably wider than that of conventional bilayers composed of QWs. The 
gapless nature of the Dirac spectrum does not favor the formation of interlayer excitons and the strong-coupling regime, $r_\mathrm{s}^{\mathrm{e}(\mathrm{h})}\gg 1$ and $p_\mathrm{F}d_\mathrm{eh}/\hbar\sim1$, does not correspond to their the Bose--Einstein condensation. Instead, the strong-coupling regime is anticipated to correspond to a multiband BCS-like paired state~\cite{EHMultiband1,EHMultiband2,EHMultiband3} in which pairing correlations also span to remote bands (filled valence bands for the Dirac electrons at the surface of the TI). For the parameters listed in the main text as well as  $\kappa_\mathrm{e}\approx 45$, $\kappa_\mathrm{e}\approx 13$, and $d_\mathrm{eh}\approx30\; \hbox{nm}$~\cite{Gamucci-Pellegrini-Natcomm-2014}, the controlling parameters listed above can be estimated as $r_\mathrm{s}^\mathrm{e}\sim 0.07$, $r_\mathrm{s}^\mathrm{h}\sim 0.36 $, and $p_\mathrm{F}d_\mathrm{eh}/\hbar\sim 3.3$. As a result, the system is in the weak- to moderate-coupling regime, which justifies the approach elaborated in the main part of this paper.

\section{Condensation energy }
The condensation energy $\mathcal E$ is given by the difference between the ground-state energies of the paired and normal states. Its density (condensation energy per unit area) is given by 
\begin{equation}
\mathcal E = \sum_\vec{p} \left( \xi_\vec{p}^\mathrm{e} n_\vec{p}^\mathrm{e} + \xi_\vec{p}^\mathrm{h} n_\vec{p}^\mathrm{h} + \varepsilon_{\vec{p}}^{-} \right) + \frac{2(|\Delta_\uparrow|^2+|\Delta_\downarrow|^2)}{V}.
\end{equation}
Here, $n_\vec{p}^{\mathrm{e}(\mathrm{h})}$ represents the occupation factors for non-interaction electrons (holes), and the integration over momentum is restricted to a ring centered at the Fermi momentum $p_\mathrm{F}$ with width $2p_0$. If we introduce $\Delta_0=2 v_0 p_0 \exp[-1/\lambda]$ and $\lambda=\nu_\mathrm{F}V/2$, the condensation energy can be presented as
\begin{equation*}
\mathcal E=\frac{\nu_\mathrm{F}}{2} \int \frac{d\phi_\vec{p}}{2\pi} \left[2 |\Delta_\vec{p}|^2 \ln \left(\frac{|\Delta_\vec{p}|}{\Delta_0}\right) -|\Delta_\uparrow|^2 - |\Delta_\downarrow|^2\right].  
\end{equation*}
The dependence of the condensation energy on the magnitude of the order parameters is presented in Fig.~2 and is discussed in the main part of the paper. 

\section{GL functional}
The GL functional for the free energy density $\mathcal F_\mathrm{GL}$ can be derived if the order parameter $\hat{\Delta}$ in the mean-field Hamiltonian, Eq.~(7), is treated in a perturbative manner. The resulting expression can be presented as
\begin{equation}
\begin{split}
\mathcal F_\mathrm{GL}=\sum_{s s'} (V^{-1}-\Lambda^a_{s s'} \Pi ) \Delta_s^* \Delta_{s'}   \\ + \frac{1}{2}\sum_{s_1 s'_1  s_2 s'_2} K  \Lambda^b_{s_1 s_1' s_2 s_2'}  \Delta_{s_1}^* \Delta_{s'_1} \Delta_{s_2}^* \Delta_{s'_2} .      
\end{split}\end{equation}
Here, the $s$-indices indicate the electron spin states $\uparrow$ and $\downarrow$. The factors $\Pi$ and $K$ are given by 
\begin{equation}
\begin{split}
\Pi=T \sum_{\omega_n,\vec{p}} \frac{1}{(i\omega_n - \xi^{\mathrm{e}}_\vec{p}) (i\omega_n - \xi^{\mathrm{h}}_\vec{p})},\\ 
K=T \sum_{\omega_n,\vec{p}} \frac{1}{(i\omega_n - \xi^{\mathrm{e}}_\vec{p})^2 (i\omega_n - \xi^{\mathrm{h}}_\vec{p})^2}.
\end{split}
\end{equation} 
Here, $\omega_n$ denotes the fermionic Matsubara frequencies $\omega_n=\pi T (2n+1)$. These terms represent the second- and fourth-order loop diagrams corresponding to a conventional electron--hole bilayer. In contrast, the helical nature of Dirac electrons is reflected in the matrix form factors, which are given by
\begin{align}
\Lambda^a_{s s'}= 2\int \frac{d\phi_\vec{p}}{2\pi} \langle \vec{p},s| \vec{p}, s'\rangle \\
\Lambda^b_{s_1 s_1' s_2 s_2'}= 4\int \frac{d\phi_\vec{p}}{2\pi} \langle \vec{p},s_1| \vec{p}, s'_1\rangle \langle \vec{p},s_2| \vec{p}, s'_2\rangle.
\end{align}
The form factor at the beginning of the quadratic term is $\Lambda^a_{s s'}=\delta_{s s'}$, which does not lead to interactions between the two order parameters. After an explicit evaluation of the second form factor $\Lambda^b_{s_1 s_1' s_2 s_2'}$, the GL functional can be presented as
\begin{eqnarray}
\mathcal F_\mathrm{GL}(|\Delta_\uparrow|,|\Delta_\downarrow |)=a ( |\Delta_\uparrow|^2+|\Delta_\downarrow|^2)\nonumber\\
    +\frac{b}{2}\left(|\Delta_\uparrow|^2+|\Delta_\downarrow|^2\right)^2 + 2 b|\Delta_\uparrow|^2|\Delta_\downarrow|^2.
\end{eqnarray}
Cumbersome explicit expressions for $a$ and $b$ are of minimal importance. The sign of the last term determines the nature of interactions between the two order parameters. Because $b>0$, the last term describes the repulsive interactions between the order parameters and causes their coexistence to be energetically unfavorable. 

\section{Domain wall states}
In the Andreev (semiclassical) approximation, the BdG equations decouple into two ($\alpha=\pm1$) independent Dirac-like equations, $\hat{K}^\alpha \chi=\varepsilon^\alpha_{p_y} \chi^\alpha$. Here, $\chi = \{\chi^\alpha_{\uparrow}, \chi^\alpha_{\downarrow}\}$ is the spinor envelope wave function, and the corresponding Hamiltonian is given by   
\begin{equation}
K^\alpha=
\begin{pmatrix} \alpha v_\mathrm{e} \cos(\bar{\phi}_\vec{p}) \hat{p}_x  & \Delta_\alpha (\bar{\phi}_\vec{p},x)\\ \Delta_\alpha^* (\bar{\phi}_\vec{p},x) & -\alpha v_\mathrm{h} \cos(\bar{\phi}_\vec{p}) \hat{p}_x \end{pmatrix}.    
\end{equation}
Here, $\bar{\phi}_\vec{p}=\arcsin(p_y/p_\mathrm{F})$ (it should be noted that its range is $[-\pi/2,\pi/2]$ while the range of the polar angle $\phi_\vec{p}$ used in the main text is $[-\pi,\pi]$), and the corresponding order parameter is given by 
\begin{equation*}
\begin{split}
\Delta_{+}(\bar{\phi}_\vec{p},x)&= \Delta_0 \left(\Theta(x) e^{i \Phi}\;e^{i\frac{\bar{\phi}_\vec{p}}{2}} \; - \;\; i \Theta(-x) \; e^{-i\frac{\bar{\phi}_\vec{p}}{2}}\right),\\
\Delta_{-}(\bar{\phi}_\vec{p},x)&= \Delta_0 \left(\Theta(x) e^{i\Phi} e^{i\frac{\pi-\bar{\phi}_\vec{p}}{2}} - i \Theta(-x) e^{-i\frac{\pi-\bar{\phi}_\vec{p}}{2}}\right).
\end{split}
\end{equation*}
The domain wall manifests solely as the phase jump, while its magnitude $|\Delta_{\alpha}(\bar{\phi}_\vec{p},x)|=\Delta_0$ is position-independent. Because of the asymmetry of the electron--hole spectrum $v_\mathrm{e}\neq v_\mathrm{h}$, the gap becomes indirect and is reduced compared with $\Delta_0$ as $\Delta_\mathrm{g}=\Delta_0 \sqrt{v_\mathrm{e} v_\mathrm{h}}/v_0$. As a result, the energy can be parameterized as $\varepsilon=\Delta_\mathrm{g}\sin(\theta)$, where $\theta$ is defined in the range $[-\pi/2,\pi/2]$. The wave functions of states localized at the domain wall are given by
\begin{equation*}
\begin{split}
\chi(x>0)=N
\begin{pmatrix}
 v_\mathrm{h} \Delta_\alpha e^{- i \alpha \theta}  \\ - i \alpha v_0
\end{pmatrix} e^{- i \alpha \tan(\theta) \frac{(v_\mathrm{e}-v_\mathrm{h}) x}{v_0 d}  - \frac{x}{d}},\\
\chi(x<0)=N
\begin{pmatrix}
- v_\mathrm{h} \Delta_\alpha e^{i \alpha \theta}  \\ - i \alpha v_0
\end{pmatrix} e^{- i \alpha \tan(\theta) \frac{(v_\mathrm{e}-v_\mathrm{h}) x}{v_0 d}   + \frac{x}{d}}.
\end{split}
\end{equation*}
Here, $N$ is the normalization coefficient, and $d=\hbar \sqrt{v_\mathrm{e} v_\mathrm{h}}/\Delta_0$ is the penetration length of the wave function for the domain wall states. The wave function is continuous across the domain wall only if the following condition is satisfied:    
\begin{equation}
\theta_{p_y}^{\alpha n}= \frac{\Phi}{2\alpha}-\frac{\pi}{4} +\bar{\phi}_\vec{p} +\pi n,  
\end{equation}
where $n$ is an integer. Only the solutions within the range $[-\pi/2,\pi/2]$ are physically relevant. As a result, the dispersion of the domain wall states is given by
\begin{equation}
\varepsilon_{p_y}^{\alpha n} =\Delta_\mathrm{g} \sin \left(\frac{\Phi}{2\alpha}-\frac{\pi}{4} + \frac{1}{2}\arcsin\left(\frac{p_y}{p_\mathrm{F}}\right) +\pi n\right).
\end{equation}
This dispersion relation is discussed and plotted in the main text of the paper.

\section{Effect of the magnetically induced gap in the Dirac spectrum}
The degeneracy between two pairing channels in the TI-QW bilayer is protected by time-reversal symmetry and can be lifted if the symmetry is broken. In the presence of an out-of-plane magnetically induced exchange field $\delta$, which plays the role of the Dirac mass (the additional term $\delta \sigma_z$ in Eq.~(1)), a gap of $2|\delta|$ is opened in the dispersion of the Dirac electrons, while the spinor wave functions are modified as  
$|\vec{p}\rangle =\{ \cos(\zeta_\vec{p})  e^{- i \phi_\vec{p}/2}, i \sin(\zeta_\vec{p}) e^{i \phi_\vec{p}/2}\}^{\mathrm{T}}$. Here, $\phi_\vec{p}$ is the polar angle for the electron momentum $\vec{p}$, and $\cos(2\zeta_\vec{p})=\delta/\sqrt{\delta^2+v_\mathrm{e}^2 p^2}$. The overlap of the spinor wave functions that shapes the Cooper pairing channels is modified as
\begin{equation}
\label{LambdaSM}
\Lambda_{\vec{p}\vec{p}'}= \cos^2(\zeta_\vec{p}) e^{i \frac{ \phi_{\vec{p}\vec{p'}}}{2}} + \sin^2(\zeta_\vec{p}) e^{-i \frac{ \phi_{\vec{p}\vec{p'}}}{2}}.
\end{equation}
As a result, the coupling constants for the two channels are $\lambda_{\uparrow(\downarrow)}=[1\pm \cos(\zeta)]\nu_\mathrm{F}V/2$, and the degeneracy between the two channels is lifted. Here, $\zeta$ is given by $\zeta_\vec{p}$ at the Fermi level $|\vec{p}|=p_\mathrm{F}$.

\begin{figure}
    \centering
    \includegraphics[scale=1.0]{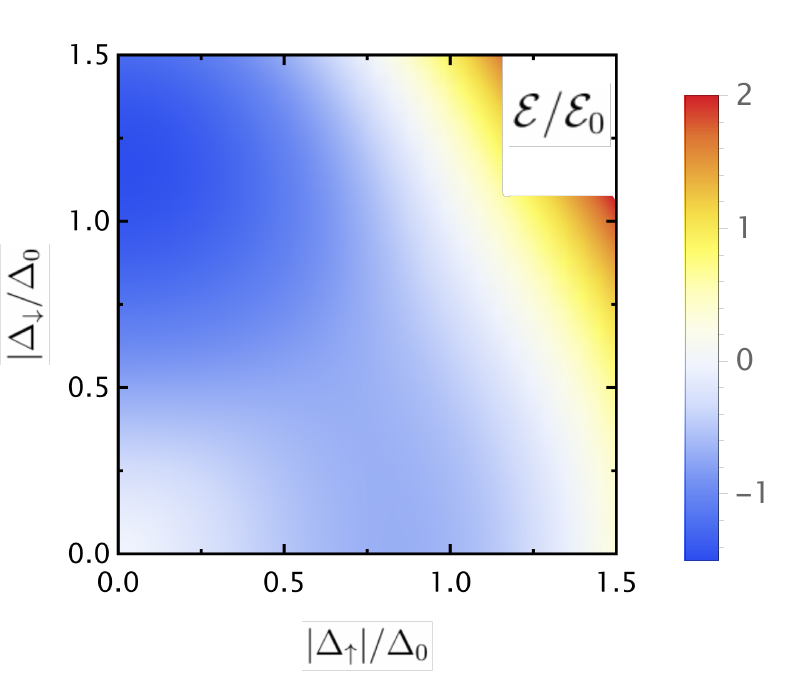}
\caption{Dependence of the condensation energy $\mathcal E$ on the magnitude of the order parameters $|\Delta_{\uparrow(\downarrow)}|$ evaluated in the presence of a small gap in the Dirac spectrum $\cos(\zeta)=0.03$. The coexistence of two order parameters is still energetically unfavorable. Instead, the degeneracy between the two paired states is lifted, and one state becomes stable whereas the other states becomes a metastable paired state.} \label{FigCondenergyAssymetry}
\end{figure}

The degeneracy lifting can be tracked in the dependence of the condensation energy on the amplitude of the order parameters $|\Delta_{\uparrow(\downarrow)}|$, as presented in Fig.~\ref{FigCondenergyAssymetry}. Even in the presence of an exchange field $\delta$, the coexistence of two order parameters is still energetically unfavorable. Instead, one states becomes the stable (or ground) state, while the other state becomes a metastable paired state. Interestingly, the stable and metastable states are swapped if the Dirac mass (the direction of the ordered magnetic moments interacting with Dirac electrons) changes its sign.

The presence of a gap of $2 |\delta|$ in the spectrum of the Dirac electrons is another source of nontrivial topologies that affect the Chern numbers. In the paired state with the order parameter $\Delta_{\uparrow(\downarrow)}$, the topological numbers for the upper Bogoliubov branch $(C^{\uparrow(\downarrow)}_{+})$, lower Bogoliubov branch $[C^{\uparrow(\downarrow)}_{-}]$, and Dirac valence band that do not participate in the Cooper pairing $[C^{\uparrow(\downarrow)}_\mathrm{Dv}]$ are given by 
\begin{align}
C^\uparrow_{+}=\frac{1}{2}, \quad C^\uparrow_{-}=-\frac{1}{2}+\frac{\delta}{2 |\delta|}, \quad C^\uparrow_{\mathrm{Dv}}=-\frac{\delta}{2 |\delta|},\\ 
C^\downarrow_{-}=-\frac{1}{2}, \quad C^\downarrow_{-}=\frac{1}{2}+\frac{\delta}{2 |\delta|}, \quad C^\downarrow_{\mathrm{Dv}}=-\frac{\delta}{2 |\delta|}. 
\end{align}
These modifications are responsible for the formation of extra chiral states within the gap separating the lower Bogoliubov branch and the Dirac valence band. However, the topological arguments for the chiral states within the gap separating the Bogoliubov branches, as discussed in the main part of the paper, remain intact.

\section{Other hybrid bilayers}
Another promising platform for hybrid Cooper pair condensates involves chiral charge carriers in few-layer graphene (e.g., graphene, Bernal-stacked bilayer graphene, or rhombohedral trilayer graphene). The low-energy electronic states in these materials are concentrated in the vicinity of two non-equivalent valleys ($\tau=\pm1$) and can be described by the following Hamiltonian:   
\begin{equation}
H(\tau,n)=
\begin{pmatrix}
    0& u^n(\tau p_x-ip_y)^n\\
    u^n(\tau p_x+ip_y)^n&0
\end{pmatrix}
\end{equation}
where $n$ is the number of layers in the few-layer graphene, and $u$ determines the dispersion slope for electrons and holes.

As argued in the main text, the presence of multiple pairing channels originates from the overlap of the spinor wave functions $\Lambda_{\vec{\vec{p},\vec{p}'}}$.
For few-layer graphene/QW bilayers \cite{EfimkinCPs}, the overlap factor does not depend on the valley index $\tau$ given by \begin{equation}
\Lambda_{\vec{\vec{p}\vec{p}'}}=\cos\left(\frac{n \phi_\vec{\vec{p}\vec{p}'}}{2}\right).
\end{equation}
As a result, this bilayer also has a pair of degenerate pairing channels (per valley), and their orbital numbers are equal to $l=\pm n/2$. 

For few-layer graphene/few-layer graphene bilayers, the factor must be substituted by 
\begin{equation*}
\bar{\Lambda}_{\vec{p}\vec{p}'}=\Lambda^{n_\mathrm{e}}_{\vec{p}\vec{p}'} \Lambda^{n_\mathrm{h}}_{\vec{p}'\vec{p}}= \cos\left(\frac{n_\mathrm{e} \phi_\vec{\vec{p}\vec{p}'}}{2}\right) \cos\left(\frac{n_\mathrm{h} \phi_\vec{\vec{p}\vec{p}'}}{2}\right). 
\end{equation*}
Here, $n_\mathrm{e}(\mathrm{h})$ indicates the number of layers in the few-layer graphene with an excess of electrons and holes. Unless $n_\mathrm{e}=n_\mathrm{h}$, the bilayer has four degenerate channels with orbital momenta $l=-n_\mathrm{e}-n_\mathrm{h}, -n_\mathrm{e}+n_\mathrm{h}, n_\mathrm{e}-n_\mathrm{h}$, and $n_\mathrm{e}+n_\mathrm{h}$. For each channel, the coupling constants are $\lambda_l=\nu_0 V/4$. For non-hybrid  bilayers $n_\mathrm{e}=n_\mathrm{h}\equiv n$, there are three pairing channels with $l=-2n, 0$, and $2n$, and the $s$-wave channel is dominant because the corresponding coupling constant $\lambda_0=\nu_0 V/2$ is two-fold larger than that for the other channels $\lambda_{\pm 2n}=\nu_0 V/4$. Thus, we conclude that the degeneracy of the pairing channels is a general feature of hybrid bilayers.

\bibliographystyle{apsrev4-1}
\bibliography{References}

\end{document}